\documentclass[showpacs,aps,12pt,graphicx]{revtex4}
\usepackage{graphicx}
\begin{document}

\title{Measurement of arbitrary two-photon entanglement state with the photonic Faraday rotation }

\author{Lan Zhou$^{1,2}$ \footnote{Email address: zhoul@njupt.edu.cn} }

\address{$^1$College of Mathematics \& Physics, Nanjing University of Posts and Telecommunications, Nanjing,
210003, China\\
 $^2$Key Lab of Broadband Wireless Communication and Sensor Network
 Technology, Nanjing University of Posts and Telecommunications, Ministry of
 Education, Nanjing, 210003,
 China\\}

\begin{abstract}
We propose an efficient protocol for measuring the concurrence of arbitrary two-photon pure entangled state with the help of the photonic Faraday rotation. In the protocol, the concurrence of the photonic entangled state can be conversed into the total success probability for picking up the odd-parity photonic state. For completing the measurement task, we require some auxiliary three-level atoms, which are trapped in the low-quality cavities. Our protocol can be well realized under current experimental conditions. Moreover, under practical imperfect atom state detection and photonic Faraday rotation conditions, our protocol can also work well. Based on these features, our protocol may be useful in current quantum information processing.
\end{abstract}
 \maketitle

\section{Introduction}
In the quantum information and computation field, entanglement plays an extremely important role. In almost all the quantum information tasks, such as the quantum teleportation \cite{teleportation,cteleportation}, quantum dense coding \cite{densecoding,densecoding1,densecoding2},
 quantum communication \cite{QSDC,QSDC1,gub1,gub2,gub3,shengpra,entropy}, quantum state
sharing \cite{QSS1,QSS2,QSS3}, and entanglement-based quantum key distribution \cite{Ekert91,QKDdeng1,QKDdeng2},
 the nonlocal parties need the entanglement to setup the quantum channel. Meanwhile, in various quantum computation tasks, such as the linear optics quantum computation \cite{optics}, one-way quantum computation \cite{oneway,zhoucluster,shengcluster}, and so on \cite{computation,computation1,computation2}, we also need to create entanglement.

 In practical applications, we usually need to know the exact information on the entanglement. In this way, during the past few years, the entanglement quantization has become an important and interesting topic in the quantum information processing. In the early works, both the entanglement witness and Bell inequality have been used to characterize the entanglement \cite{witness,witness1,witness2,bell}. However, they can only disclose the entanglement of some quantum states but fail for other states. In this way, they cannot provide satisfactory results in general. In 1999, White \emph{et al.} proposed a straightforward method for measuring the entanglement based on the complete tomographic reconstruction of the quantum state \cite{measurement1}. However, the reconstruction of a two-qubit state requires to know 15 parameters, which is so complicated in practical operations. So far, the entanglement of formation (EOF), which is firstly proposed by Bennett \emph{et al.} in 1996, has been the most valuable entanglement quantization method \cite{EOF,EOF1}. The EOF is often used to quantify the minimal cost to prepare a certain quantum state. It has been shown that the EOF for an arbitrary two-qubit state can be exactly defined by the concurrence (C). For an arbitrary two-qubit state $\rho$, the concurrence can be defined as \cite{EOF1,mix,concurrence}
\begin{eqnarray}
C(\rho)=max\{0,\lambda_{1}-\lambda_{2}-\lambda_{3}-\lambda_{4}\},
\end{eqnarray}
where $\lambda_{i}$ (i=1,2,3,4) are the non-negative eigenvalues in decreasing order of the Hermitian matrix $R=\sqrt{\sqrt{\rho}\tilde{\rho}\sqrt{\rho}}$. $\tilde{\rho}=(\sigma_{y}\otimes\sigma_{y})\rho^{*}(\sigma_{y}\otimes\sigma_{y})$, where $\sigma_{y}$ and $\rho^{*}$ are the usual Pauli operator and the complex conjugate of $\rho$. For a pure state $|\Psi\rangle$, the concurrence can be rewritten as \cite{mix,wootter2}
\begin{eqnarray}
C(|\Psi\rangle)=|\langle\Psi^{*}|\sigma_{y}\otimes\sigma_{y}|\Psi\rangle|,
 \sigma_{y}=
 \left[
 \begin{array}{cc}
 0 &-i\\
 i& 0
 \end{array}
 \right].
\end{eqnarray}
where $\sigma_{y}$ is the Pauli matrix of $|\Psi\rangle$. Moreprecisely, for a general two-qubit pure state as $|\Psi\rangle=a_{1}|00\rangle+a_{2}|01\rangle+a_{3}|10\rangle+a_{4}|11\rangle$, where $|a_{0}|^{2}+|a_{1}|^{2}+|a_{2}|^{2}+|a_{3}|^{2}=1$, the concurrence can be simplified as
\begin{eqnarray}
C(|\Psi\rangle)=2|a_{1}a_{4}-a_{2}a_{3}|.
\end{eqnarray}

In 2006, Walborn \emph{et al.} experimentally realized the determination of entanglement with a single
measurement in linear optics \cite{concurrence4}. In their experiment, they successfully measured the concurrence for a partially entangled state $|\varphi\rangle=\alpha|01\rangle+\beta|10\rangle$, whose concurrence can be written as $C=2|\alpha\beta|$. Later, Romero \emph{et al.} proposed an effective way for detecting the concurrence of pure atomic state with the help of the controlled-not (CNOT) gate \cite{concurrence3}. Recently, the group of Cao put forward two efficient methods for measuring the concurrence of the two-photon polarization entangled pure or mixed states, with the help of the cross-Kerr nonlinearity \cite{concurrence,concurrence1}.

During the past few year, available techniques have achieved the input-output process relevant to optical cavities with low-quality (Q) factors, such
as the microtoroidal resonator (MTR) \cite{MTR}. It is attractive and applicable to combine the input-output
process with low-Q cavities, for if achieved, it can accomplish high-Q quantum tasks with currently available techniques. In 2009, An \emph{et al.} put forward an innovative scheme to implement the quantum information tasks by moderate cavity-atom coupling with low-Q cavities \cite{r}. They have shown that when the photon interacts with an atom trapped in a low-Q cavity, differently polarized photons can make different phase rotation, which is called photonic Faraday rotation. The photonic Faraday rotation only works in low-Q cavities and is insensitive to both cavity decay and atomic spontaneous emission. In 2012, based on the photonic Faraday rotation, the group of Peng proposed two entanglement concentration protocols (ECP) for arbitrary atomic state and photonic states, respectively \cite{Pengpra}, which were improved by our group later \cite{zhouatom,zhouphoton,shengqip1,shengqip2}. In 2008, Lee \emph{et al.} adopted the atoms as the flying qubits to realize the measurement for the concurrence in a cavity quantum electrodynamics (QED) system \cite{concurrence2}.

Inspired by the previous works, in the paper, we design a protocol for measuring the concurrence of arbitrary two-photon pure entangled state with the photonic Faraday rotation in a low-Q cavity QED system. We require some auxiliary three-level atoms, which are trapped in the low-Q cavities. We make the photon pulse pass through the low-Q cavity and react with the three-level atom. Taking use of the photonic Faraday rotation, we can converse the concurrence of the photonic state into the total success probability for picking up the odd-parity photonic state, which can be measured directly.  Our protocol does not require the complicated operation and can be realized under current imperfect experimental conditions. In this way, our protocol may be useful in current quantum information processing.

\section{The basic principle of photonic Faraday rotation}
\begin{figure}[!h]
\begin{center}
\includegraphics[width=6cm,angle=0]{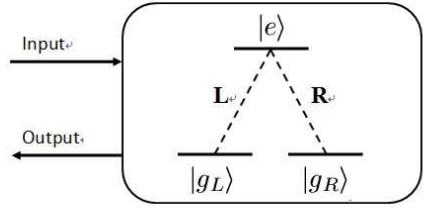}
\caption{A schematic drawing of the interaction between the photon pulse and the three-level atom in the low-Q cavity. We make a three-level atom  trap in a low-Q cavity. $|g_{L}\rangle$ and $|g_{R}\rangle$ represent the two Zeeman sublevels of its degenerate ground state, and $|e\rangle$ represents its excited state. The state $|g_{L}\rangle$ and $|g_{R}\rangle$ couple with a left (L) polarized and a right (R) polarized photon, respectively.}
\end{center}
\end{figure}

The photonic Faraday rotation is the key operation in our protocol. We would introduce the basic principle of the photonic Faraday rotation, before explaining the details of our protocol. As shown in Fig. 1, suppose that a three-level atom is trapped in the one-side low-Q cavity. The states $|g_{L}\rangle$ and $|g_{R}\rangle$ represent the two Zeeman sublevels of its degenerate ground state, and $|e\rangle$ represents its excited state. A single photon pulse with the form of $|\varphi_{in}\rangle=\frac{1}{\sqrt{2}}(|L\rangle+|R\rangle)$ enters the cavity and reacts with the three-level atom, where $|L\rangle$ and $|R\rangle$ represent the left-circularly polarization and right-circularly polarization of the input photon, respectively. When the atom absorbs or emits a $|L\rangle$ ($|R\rangle$) polarized photon, it will cause the transitions $|g_{L}\rangle\leftrightarrow|e\rangle$ ($|g_{R}\rangle\leftrightarrow|e\rangle$).

  The Hamiltonian of the whole system can be written as \cite{r,fengmang3,cteleportationqip,Faraday},
\begin{eqnarray}
H=H_{0}+\hbar\lambda\sum_{j=L,R}(a^{\dagger}_{j}\sigma_{j-}+a_{j}\sigma_{j+})+H_{R},\label{hamiltonian}
\end{eqnarray}
with
\begin{eqnarray}
H_{0}=\sum_{j=L,R}[\frac{\hbar\omega_{0}}{2}\sigma_{jz}+\hbar\omega_{c}a^{\dagger}_{j}a_{j}],\label{h0}
\end{eqnarray}
and
\begin{eqnarray}
H_{R}&=&H_{R0}+i\hbar[\int_{-\infty}^{\infty}d\omega\sum_{j=L,R}\alpha(\omega)(b^{\dagger}_{j}(\omega)a_{j}+b_{j}(\omega)a^{\dagger}_{j})\nonumber\\
&+&\int_{-\infty}^{\infty}d\omega\sum_{j=L,R}\bar{\alpha}(\omega)(c^{\dagger}_{j}(\omega)\sigma_{j-}+c_{j}(\omega)\sigma_{j+})].\label{hr}
\end{eqnarray}
Here, $\lambda$ is the atom-field coupling constant. $a^{\dagger}_{j}$ and $a_{j}$ ($j=L, R$) are the creation and annihilation operators of the filed-mode in the cavity, respectively. $\sigma_{L-}$ and $\sigma_{L+}$ ($\sigma_{R-}$ and $\sigma_{R+}$) are the lowering and raising operators of the transition L (R), respectively. $\omega_{c}$, $\omega_{p}$, and $\omega_{0}$ represent the field frequency, photon frequency and atomic frequency, respectively. $H_{R0}$ represents the Hamiltonian of the free reservoirs, and $b_{j}$, $c_{j}$ ($b^{\dagger}_{j}$ and $c^{\dagger}_{j}$) are the annihilation (creation) operators of the reservoirs.

By solving the Langevin equations of
motion for cavity and atomic lowering operators analytically, we can obtain a single relation between the input and output single-photon state as \cite{r}
\begin{eqnarray}
r(\omega_{p})\equiv\frac{a_{out,j(t)}}{a_{in,j(t)}}=\frac{[i(\omega_{c}-\omega_{p})-\frac{\kappa}{2}][i(\omega_{0}-\omega_{p})+\frac{\gamma}{2}]+\lambda^{2}}
{[i(\omega_{c}-\omega_{p})+\frac{\kappa}{2}][i(\omega_{0}-\omega_{p})+\frac{\gamma}{2}]+\lambda^{2}}.\label{r}
\end{eqnarray}
Here, $\kappa$ and $\gamma$ are the cavity damping rate and atomic decay
rate, respectively. If the atom uncouples to the cavity, which makes $\lambda=0$, we can simplify Eq. (\ref{r}) as
\begin{eqnarray}
r_{0}(\omega_{p})=\frac{i(\omega_{c}-\omega_{p})-\frac{\kappa}{2}}{i(\omega_{c}-\omega_{p})+\frac{\kappa}{2}}.\label{r0}
\end{eqnarray}
Eq. (\ref{r0}) can be written as a pure phase shift $r_{0}(\omega_{p})=e^{i\phi_{0}}$. On the other hand, in the interaction process, as the photon experiences an extremely weak absorption, we can consider that the output photon only experiences a pure phase shift without any absorption. In this way, with strong $\kappa$, weak $\gamma$ and $g$, Eq. (\ref{r}) can be simplified to $r(\omega_{p})\simeq e^{i\phi}$. Therefore, if the photon pulse takes action, the output photon state will convert to $|\varphi_{out}\rangle=r(\omega_{p})|L (R)\rangle\simeq e^{i\phi}|L (R)\rangle$, otherwise, the single-photon would only sense the empty cavity, and the output photon state will convert to $|\varphi_{out}\rangle=r_{0}(\omega_{p})|L (R)\rangle= e^{i\phi_{0}}|L (R)\rangle$.

In this way, for an input single-photon state $|\varphi_{in}\rangle=\frac{1}{\sqrt{2}}(|L\rangle+|R\rangle)$, if the initial atom state is $|g_{L}\rangle$, the output photon state can evolve to
\begin{eqnarray}
|\varphi_{out}\rangle_{-}=\frac{1}{\sqrt{2}}(e^{i\phi}|L\rangle+e^{i\phi_{0}}|R\rangle),\label{L}
\end{eqnarray}
while if the initial atom state is $|g_{R}\rangle$, the output photon state will evolve to
\begin{eqnarray}
|\varphi_{out}\rangle_{+}=\frac{1}{\sqrt{2}}(e^{i \phi_{0}}|L\rangle+e^{i \phi}|R\rangle).\label{R}
\end{eqnarray}
Based on Eq. (\ref{L}) and Eq. (\ref{R}), the angle $\Theta^{-}_{F}=\phi_{0}-\phi$ or $\Theta^{+}_{F}=\phi-\phi_{0}$ is defined as the photonic Faraday rotation.

In  Eq. (\ref{L}) and Eq. (\ref{R}), it can be found that in a certain case, i.e., $\omega_{0}=\omega_{c}$, $\omega_{p}=\omega_{c}-\frac{\kappa}{2}$, and $g=\frac{\kappa}{2}$, we can obtain $\phi=\pi$ and $\phi_{0}=\frac{\pi}{2}$, so that the relation between the input and output photonic state can be written  as \cite{Pengpra}
\begin{eqnarray}
&&|L\rangle|g_{L}\rangle\rightarrow -|L\rangle|g_{L}\rangle,\qquad |R\rangle|g_{L}\rangle\rightarrow i|R\rangle|g_{L}\rangle,\nonumber\\
&&|L\rangle|g_{R}\rangle\rightarrow i|L\rangle|g_{R}\rangle,\qquad |R\rangle|g_{R}\rangle\rightarrow -|R\rangle|g_{R}\rangle.\label{rule}
\end{eqnarray}

\section{Detection of the concurrence of arbitrary two-photon entangled state}
\begin{figure}[!h]
\begin{center}
\includegraphics[width=7cm,angle=0]{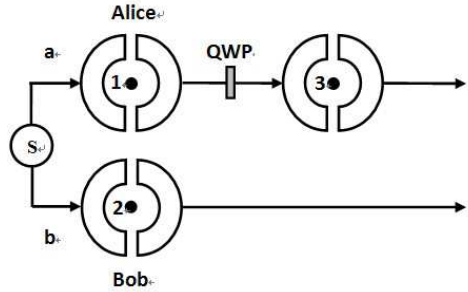}
\caption{A schematic drawing of the principle of our detection protocol. Three three-level atoms, here named "1", "2", and "3" are trapped in three low-Q cavities, respectively. All the three atoms have the same quantum state as $\frac{1}{\sqrt{2}}(|g_{L}\rangle+|g_{R}\rangle)$. Alice and Bob share two pairs of photonic states with the form of $|\phi\rangle=\alpha|RR\rangle_{ab}+\beta|RL\rangle_{ab}+\gamma|LR\rangle_{ab}+\delta|LL\rangle_{ab}$, which are generated by the single photon source $S$. The QWP represents the quarter wave plate.}
\end{center}
\end{figure}

Now, we begin to introduce our protocol for detecting the concurrence of the photonic entangled state with the help of the photonic Faraday rotation. We suppose two nonlocal parties, Alice and Bob, share two pairs of arbitrary two-photon entangled states with the same form of
\begin{eqnarray}
|\phi_{1}\rangle_{a1b1}&=&\alpha|RR\rangle_{a1b1}+\beta|RL\rangle_{a1b1}+\gamma|LR\rangle_{a1b1}+\delta|LL\rangle_{a1b1},\nonumber\\
|\phi_{2}\rangle_{a2b2}&=&\alpha|RR\rangle_{a2b2}+\beta|RL\rangle_{a2b2}+\gamma|LR\rangle_{a2b2}+\delta|LL\rangle_{a2b2},
\end{eqnarray}
where $\alpha$, $\beta$, $\gamma$, and $\delta$ are the entanglement coefficients of the photonic state, and $|\alpha|^{2}+|\beta|^{2}+|\gamma|^{2}+|\delta|^{2}=1$.

In this way, the whole four-photon state can be written as
\begin{eqnarray}
&&|\Phi\rangle_{a1a2b1b2}=|\phi_{1}\rangle_{a1b1}\otimes|\phi_{2}\rangle_{a2b2}\nonumber\\
&=&(\alpha^{2}|RRRR\rangle+\alpha\beta|RRRL\rangle+\alpha\gamma|RLRR\rangle+\alpha\delta|RLRL\rangle\nonumber\\
&+&\alpha\beta|RRLR\rangle+\beta^{2}|RRLL\rangle+\beta\gamma|RLLR\rangle+\beta\delta|RLLL\rangle\nonumber\\
&+&\alpha\gamma|LRRR\rangle+\beta\gamma|LRRL\rangle+\gamma^{2}|LLRR\rangle+\gamma\delta|LLRL\rangle\nonumber\\
&+&\alpha\delta|LRLR\rangle+\beta\delta|LRLL\rangle+\gamma\delta|LLLR\rangle+\delta^{2}|LLLL\rangle)_{a1a2b1b2}.\label{whole}
\end{eqnarray}

As shown in Fig. 2, Alice and Bob prepare three three-level atoms, here named $"1"$, $"2"$, and $"3"$. All the three atoms are in the same state as $|\varphi\rangle=\frac{1}{\sqrt{2}}(|g_{L}\rangle+|g_{R}\rangle)$, and trapped in three low-Q cavities, respectively. Each of two parties makes the two photons in his/her hand pass through one low-Q cavity and interact with the atom, successively.  Based on Eq. (\ref{rule}), we can obtain the relation between the input and output photon-atom state as
\begin{eqnarray}
&&|R\rangle|R\rangle(|g_{L}\rangle+|g_{R}\rangle)\rightarrow|R\rangle|R\rangle(-|g_{L}\rangle+|g_{R}\rangle),\nonumber\\
&&|L\rangle|L\rangle(|g_{L}\rangle+|g_{R}\rangle)\rightarrow|L\rangle|L\rangle(|g_{L}\rangle-|g_{R}\rangle),\nonumber\\
&&|R\rangle|L\rangle(|g_{L}\rangle+|g_{R}\rangle)\rightarrow -i|R\rangle|L\rangle(|g_{L}\rangle+|g_{R}\rangle),\nonumber\\
&&|L\rangle|R\rangle(|g_{L}\rangle+|g_{R}\rangle)\rightarrow -i|L\rangle|R\rangle(|g_{L}\rangle+|g_{R}\rangle).\label{rule2}
\end{eqnarray}

It can be found that if the two photons are in the even parity, such as $|RR\rangle$ or $|LL\rangle$, the atomic state will be changed, while if they are in the odd parity, such as $|RL\rangle$ or $|LR\rangle$, the atomic state will keep the same. After the photon-atom interaction, Alice and Bob detect the atomic states. If they select the items which make the atomic states not change, Eq. (\ref{whole}) will collapse to
\begin{eqnarray}
|\Phi_{1}\rangle_{a1a2b1b2}&=&\alpha\delta(|RLRL\rangle_{a1a2b1b2}+|LRLR\rangle_{a1a2b1b2})\nonumber\\
&+&\beta\gamma(|RLLR\rangle_{a1a2b1b2}+|LRRL\rangle_{a1a2b1b2}),\label{select}
\end{eqnarray}
with the success probability of $P_{1}=2|\alpha\delta|^{2}+2|\beta\gamma|^{2}$.

Next, Alice performs the Hadamard (H) operation on the photons in $a1$ and $a2$ modes by making them pass through the quarter wave plate (QWP). The H operation can make
\begin{eqnarray}
|R\rangle\rightarrow\frac{1}{\sqrt{2}}(|R\rangle+|L\rangle),\qquad
|L\rangle\rightarrow\frac{1}{\sqrt{2}}(|R\rangle-|L\rangle).
\end{eqnarray}
After that, Eq. (\ref{select}) will evolve to
\begin{eqnarray}
|\Phi_{2}\rangle_{a1a2b1b2}&=&(\alpha\delta+\beta\gamma)(|RR\rangle-|LL\rangle)_{a1a2}(|RL\rangle+|LR\rangle)_{b1b2}\nonumber\\
&+&(\alpha\delta-\beta\gamma)(|LR\rangle-|RL\rangle)_{a1a2}(|RL\rangle-|LR\rangle)_{b1b2}.\label{QWP}
\end{eqnarray}

 Then, Alice makes the two photons in a1 and a2 modes enter another cavity and interact with the atom "3". According to Eq. (\ref{rule2}), Alice also selects the items, which make the atomic state not change. In this way, Eq. (\ref{QWP}) will collapse to
\begin{eqnarray}
|\Phi_{3}\rangle_{a1a2b1b2}=(|LR\rangle-|RL\rangle)_{a1a2}(|RL\rangle-|LR\rangle)_{b1b2},\label{result}
\end{eqnarray}
with the success probability of $P_{2}=\frac{|\alpha\delta-\beta\gamma|^{2}}{2(|\alpha\delta|^{2}+|\beta\gamma|^{2})}$.

Therefore, the total success probability for obtaining the state in Eq. (\ref{result}) can be calculated as
\begin{eqnarray}
P_{total}=P_{1}P_{2}=|\alpha\delta-\beta\gamma|^{2}=\frac{C^{2}(\phi)}{4}.\label{relation}
\end{eqnarray}
Based on the relationship between the concurrence and the total success probability for obtaining Eq. (\ref{result}), we can obtain the concurrence of the photonic state by detecting the $P_{total}$. Of course, in the practical experiment, in order to measure the $P_{total}$ exactly, we need to repeat the protocol for a large number of times and consume large number of photonic and atomic states.

\section{Discussion and summary}
In the paper, we propose an effective protocol for detecting the concurrence of arbitrary two-photon pure entangled state. In the protocol, we adopt the photonic Faraday rotation to construct the quantum nondemolition measurement for the photonic states. After the photon-atom interaction, by measuring the atomic state in the low-Q cavity and selecting the photonic items which make the atomic state not change, we can distill the photonic states in the odd parity, with the success probability of $P_{total}$. Based on the relationship between $P_{total}$ and the concurrence in Eq. (\ref{relation}), we can directly obtain the concurrence of the arbitrary two-photon entangled state by exactly calculating the $P_{total}$.
In our protocol, in order to realize the photonic Faraday rotation, the three-level atom encoded in the low-Q cavity is the key element. Under current experimental conditions, $^{87}Rb$ and $^{85}Rb$ atoms encoded in the fiber-based Fabry-Perot cavity have been proved to be good candidate systems. In Ref. \cite{colombe}, they choose the states of $|F=2\rangle$, $m_{F}=\pm1$ of the $5S_{1/2}$ to be the two ground states $|g_{L}\rangle$ and $|g_{R}\rangle$, respectively. Under this case, the transition frequency between the ground states and the excited state at $\lambda=780 nm$ is $\omega_{0}=\frac{2 \pi c}{\lambda} \approx 2.42 \times 10^{15} Hz$. The cavity length, cavity rate and the finesse are $L=38.6 \mu m$, $\kappa=2\pi \times 53 MHz$ and F=37000, respectively. In the past few years, certain studies on the atom state detection of $^{85}Rb$ atoms coupled to an optical cavity have also been reported. For example, in 2005, Nu$\beta$mann \emph{et al.} have successfully controlled and adjusted the individual ultracold $^{85}Rb$ atoms coupled to a high-finesse optical cavity \cite{numann}. In 2007, Fortier \emph{et al.} realized the deterministic loading of single $^{85}Rb$ atoms in a cavity by incorporating a deterministically loaded atom conveyor \cite{fortier}. Based on the early experimental works, our protocol can be well realized under current experimental conditions.

In the above description, our protocol is operated under the ideal conditions that the atom detection efficiency $\eta_{a}=100\%$ and the photonic Faraday rotation angle $\Theta^{+}_{F}=\phi-\phi_{0}=\frac{\pi}{2}$. Actually, in practical experiment, both the imperfect detection of the atom state and the imperfect photonic Faraday rotation angle will cause some influence on the protocol.

In our protocol, for completing the detection task, we need to detect the atom state for three times. In experiment, the atom state detection efficiency $\eta_{a}<100\%$. Considering this imperfect detection, the description of $P_{total}$ can be revised as
\begin{eqnarray}
P'_{total}=\eta^{3}_{a}|\alpha\delta-\beta\gamma|^{2}=\eta^{3}_{a}\frac{C^{2}(\phi)}{4}.\label{relation1}
\end{eqnarray}
In this way, in practical experiment, we need to detect $\eta_{a}$ first. In the past few years, certain studies on the atom state detection of $^{85}Rb$ atoms coupled to an optical cavity have been reported. For example, Heine \emph{et al.} reported their research result on the single atom detection. They have achieved $\eta_{a}= 66\%$ in the experiment \cite{atomdetection}. Moreover, they have shown that with some improvement, the single atom detection efficiency can achieve $\eta_{a}> 95\%$ in theory.

On the other hand, in practical experiment, it is quite difficult to exactly control $\Theta^{+}_{F}=\pi/2$ exactly. Therefore, it is quite possible that $\Theta^{+}_{F}=\frac{\pi}{2}+\sigma$, where $\sigma$ is a small quantity. Under this case, Eq. (\ref{rule2}) can be modified as
\begin{eqnarray}
&&|R\rangle|R\rangle(|g_{L}\rangle+|g_{R}\rangle)\rightarrow|R\rangle|R\rangle e^{2i\phi_{0}}(|g_{L}\rangle-e^{2i\sigma}|g_{R}\rangle),\nonumber\\
&&|L\rangle|L\rangle(|g_{L}\rangle+|g_{R}\rangle)\rightarrow|L\rangle|L\rangle e^{2i\phi_{0}}(-e^{2i\sigma}|g_{L}\rangle+|g_{R}\rangle),\nonumber\\
&&|R\rangle|L\rangle(|g_{L}\rangle+|g_{R}\rangle)\rightarrow |R\rangle|L\rangle e^{i(\phi+\phi_{0})}(|g_{L}\rangle+|g_{R}\rangle),\nonumber\\
&&|L\rangle|R\rangle(|g_{L}\rangle+|g_{R}\rangle)\rightarrow |L\rangle|R\rangle e^{i(\phi+\phi_{0})}(|g_{L}\rangle+|g_{R}\rangle).\label{rule4}
\end{eqnarray}

It can be found that if the two input photons in the odd parity as $|R\rangle|L\rangle$ and $|L\rangle|R\rangle$, the atomic state will also not change. However, if the two input photons are in the even parity as $|R\rangle|R\rangle$ or $|L\rangle|L\rangle$, the atomic state will change to $|g_{L}\rangle-e^{2i\sigma}|g_{R}\rangle$ or $|g_{R}\rangle-e^{2i\sigma}|g_{L}\rangle$. This case may also contribute to the  $\frac{1}{\sqrt{2}}(|g_{L}\rangle+|g_{R}\rangle)$ with some probability and cause the detection error. Therefore, in each parity check process, we can calculate the error probability as
\begin{eqnarray}
P_{e}=|\frac{1}{\sqrt{2}}(|g_{L}\rangle+|g_{R}\rangle)\frac{1}{\sqrt{2}}(-e^{2i\sigma}|g_{L}\rangle+|g_{R}\rangle)|^{2}
=\frac{|-e^{i2\sigma}+1|^{2}}{4}.
\end{eqnarray}
In this way, in the practical experiment, the detected success probability in each parity check process can be modified as
\begin{eqnarray}
P'_{1}&=&P_{1}+(1-P_{1})\frac{|-e^{i2\sigma}+1|^{2}}{4},\nonumber\\
P'_{2}&=&P_{2}+(1-P_{2})\frac{|-e^{i2\sigma}+1|^{2}}{4}.\label{PP}
\end{eqnarray}

Based on Eq. (\ref{PP}), if the value of $\sigma$ is known, we can easily calculate exact value of $P_{1}$ and $P_{2}$, and calculate the exact value of $P_{total}$. Therefore, under this imperfect photonic Faraday rotation condition, we can also obtain the concurrence of arbitrary photonic entanglement state.

In summary, in the paper, we put forward an efficient protocol for detecting the concurrence of arbitrary two-photon entangled state. In the protocol, we adopt the photonic Faraday rotation to construct the quantum nondemolition measurement for the photonic state and the concurrence can be conversed into the total success probability for picking up the odd-parity photonic state. For completing the detection task, we need to consume large number of photonic states and auxiliary three-level atoms, each of which is trapped in a low-Q cavity.
 Comparing with other detection protocol, we do not require the sophisticated CNOT gate or single-photon detector, and our protocol is feasible under current experimental technique. Moreover, under imperfect atom detection and photonic Faraday rotation angle, our protocol can also be well realized. All the features make our protocol useful in current quantum information processing.

\section{Acknowledgements}
This work is supported by the National Natural Science Foundation of
China under Grant Nos. 11104159 and 11347110, the open research funds of Key Lab of Broadband Wireless Communication and Sensor Network Technology, Nanjing University of Posts and Telecommunications, Ministry of Education (No. NYKL201303), and the Priority Academic Program Development of Jiangsu
Higher Education Institutions.

\end{document}